\documentclass [12pt]{article}

\newcommand{\keywords}[1]{\par\parindent 0pt KEYWORDS: #1}
\newcommand{\qed}{\par\hfill$\bf\Omega$}

\newtheorem {lemma}{Lemma}
\newtheorem {proposition}{Proposition}
\newtheorem {definition}{Definition}
\newtheorem {example}{Example}
\newtheorem {notation}{Notation}
\newtheorem {corollary}{Corollary}
\newtheorem {theorem}{Theorem}
\newenvironment {acknowledgement}
  {\par\parindent 0pt Acknowledgement\par}
  {}
\newenvironment {proof}
  {\par\parindent 0pt Proof\par}
  {}

\begin {document}

\title {$C^*$-Multipliers, crossed product algebras, and\\
        canonical commutation relations}

\author{Jan Naudts\\
\small
Departement Natuurkunde, Universiteit Antwerpen UIA,\\
\small
Universiteitsplein 1, B2610 Antwerpen, Belgium.\\
\small
E-mail: naudts@uia.ua.ac.be
}

\date{February 2000}

\maketitle
\begin {abstract}

The notion of a multiplier of a group $X$ is generalized to that of a 
$C^*$-multiplier by allowing it to have values in an arbitrary 
$C^*$-algebra $\cal A$. On the other hand, the notion of the action of 
$X$ in $\cal A$ is generalized to that of a projective action of $X$ as 
linear transformations of the space of continuous functions with compact 
support in $X$ and with values in $\cal A$. It is shown that there 
exists a one-to-one correspondence between $C^*$-multipliers and 
projective actions. $C^*$-multipliers have been used to define twisted 
group algebras. On the other hand, the projective action $\tau$ can be 
used to construct the crossed product algebra ${\cal A}\times_\tau X$. 
Both constructions are unified in the present approach.

The results are applicable in mathematical physics. The multiplier 
algebra of the crossed product algebra ${\cal A}\times_\tau X$ contains 
Weyl operators $\{W(x),x\in X\}$. They satisfy canonical commutation 
relations w.r.t.~the $C^*$-multiplier. Quantum spacetime is discussed as 
an example.

\keywords {Crossed product algebra, Twisted group algebra, Multipliers,
Canonical commutation relations, Quantum spacetime.}
\end {abstract}

%%%%%%%%%%%%%%%%%%%%%%%%%%%%%%%%%%%%%%%%%%%%%%%%%%%%%%%%%%%%%%%%%%%%%%%%%%%
\section {Introduction}

Starting point for this work is the observation that the construction of 
the crossed product of a $C^*$-algebra with a group is very similar to 
the construction of twisted group algebras and of the $C^*$-algebra of 
canonical commutation relations (CCR). The convolution product of the 
group algebra ${\cal L}_1(X)$ of a locally compact group $X$ can be 
deformed in two ways. In the first case the integrable functions of $X$ 
are allowed to have values in a $C^*$-algebra  $\cal A$, and the 
deformation involves a representation $\tau$ of $X$ as homomorphisms of 
$\cal A$. This is the basis for the definition of the crossed product 
algebra ${\cal A}\times_\tau X$. In the other case a multiplier 
$\xi:\,X\times X\rightarrow {\bf C}$ is used, which leads to the notion 
of a twisted group algebra. By a slight generalization both constructs 
can be unified. The generalization of the crossed product algebra is 
obtained by replacing the representation $\tau$ by a projective action. 
The generalization of the twisted group algebra is obtained by allowing 
the multiplier to have values in an arbitrary $C^*$-algebra $\cal A$.

A short history of the crossed product algebra can be found in the 
introduction of \cite {TM67}. Its physical importance was brought out in 
\cite {DKR66}. The twisted group algebra was studied in \cite {EL69} and 
\cite {BS70}. A special case of twisted group algebra, of relevance in 
mathematical physics, is the algebra of CCR, introduced in \cite {MJ68}, 
and, independently, in \cite {SJ72}. An introductory treatment is found 
in \cite {PD90}.

An important generalization, orthogonal to the present one, is obtained
by replacing the group $X$ by a groupoid. See \cite {CA79}, or \cite
{RJ80}, Ch.~II, Sec.~1. The combination of both generalizations is not
considered here.

The structure of the paper is as follows. In the next section the main 
results are stated. Then theorem 1 is proved. In section \ref {constr} 
the crossed product algebra is constructed. Theorem 2 is proved in 
section \ref {thm2sect}. The final section discusses quantum spacetime 
as an application in quantum mechanics.

%%%%%%%%%%%%%%%%%%%%%%%%%%%%%%%%%%%%%%%%%%%%%%%%%%%%%%%%%%%%%%%%%%%%%%%%%%%
\section {Main results}

Recall that a multiplier $\xi$ of a group $X$ is a map
$\xi:\,X\times X\rightarrow{\bf C}_1\equiv
\{\alpha\in {\bf C}:\,|\alpha|=1\}$
satisfying $\xi(x,e)=\xi(e,y)=1$ for all $x\in X$, and
satisfying the {\sl cocycle property}
\begin {equation}
\xi(x,y)\xi(xy,z)=\xi(x,yz)\xi(y,z),
\qquad x,y,z\in X
\label {cocycle}
\end {equation}
($e$ is the unit element of $X$).
The notion of multiplier is generalized as follows (see \cite {BS70}).

\begin {definition}
A $C^*$-{\sl multiplier} of a group $X$ acting in a $C^*$-algebra
$\cal A$ is a map $\xi$ of $X\times X$ into the unitary elements of the
multiplier algebra $M(\cal A)$ of $\cal A$ satisfying the following
axioms.
\begin {description}
\item {(M1)} $\xi(x,e)=\xi(e,y)=1$ for all $x,y\in X$.
\item {(M2)} There exists a map $\sigma$ of $X$ into the automorphisms of
             $\cal A$ such that $\sigma_e$ is the identity transformation
             and one has
             \begin {equation}
	            \sigma_x\xi(y,z)=\xi(x,y)\xi(xy,z)\xi(x,yz)^*,
             \qquad x,y,z\in X
             \label {sigmadef}
             \end {equation}
\item {(M3)} $\sigma$ satisfies
             \begin {equation}
             \sigma_x\sigma_ya=\xi(x,y)(\sigma_{xy}a)\xi(x,y)^*,
             \qquad x,y\in X,a\in {\cal A}
             \label {projtrans}
             \end {equation}
\item {(M4)} The map $y\in X \rightarrow a\xi(x,y)$
             is continuous for all $x\in X$ and $a\in {\cal A}$.
             The maps $x\in X\rightarrow a\xi(x,x^{-1})$
             and $x\in X\rightarrow \sigma_xa$ are continuous
             for all $a\in {\cal A}$.
\end {description}
$\xi$ is also called an ${\cal A}$-multiplier of $X$.
\end {definition}

\noindent
If ${\cal A}={\bf C}$ then $\sigma$ is trivial and (\ref {sigmadef})
reduces to (\ref {cocycle}).
If the *-algebra generated by the range of $\xi$
is dense in $M({\cal A})$ then
$\sigma$ is uniquely determined by (\ref {sigmadef}).
In that case (M3) follows from (M2).
In general, $\sigma$ is not a group representation, but is
twisted by means of the $C^*$-multiplier, according to (\ref {projtrans}).
For convenience, the map $\sigma$ is
called a {\sl twisted representation} associated with $\xi$.
In \cite {BS70} the pair $(\xi,\sigma)$ is called a twisting pair
for $X$ and $\cal A$ and is considered in a more general setting,
with $\cal A$ an involutive Banach algebra, and with the
continuity requirements (M4) replaced by measurability conditions.

\begin {notation}
Let ${\cal C}_c(X)$ denote the space of complex continuous functions
with compact support in $X$. Similarly, let ${\cal C}_c(X,{\cal A})$
denote the space of continuous functions with compact support in $X$ and
values in the $C^*$-algebra $\cal A$. 
\end {notation}

Given a $C^*$-multiplier $\xi$ and an associated twisted
representation $\sigma$,
one can define a map $\tau$ of $X$ into the linear transformations of
${\cal C}_c(X,{\cal A})$ by
\begin {equation}
\tau_xf(y)=(\sigma_xf(x^{-1}y))\xi(x,x^{-1}y),
\qquad x,y\in X,f\in {\cal C}_c(X,{\cal A})
\label {taudef}
\end {equation}
It is straightforward to verify that $\tau$ satisfies the axioms
of the following definition.

\begin{definition}
Let $X$ be a locally compact Hausdorff group and let $\cal A$ be a $C^*$-algebra.
A map $\tau$ of $X$ into the linear transformations of
${\cal C}_c(X,{\cal A})$ is a {\sl projective action} if it satisfies
\begin{description}
\item {(A1)} \label {identity}
             $\tau_e$ is the identity transformation of
             ${\cal C}_c(X,{\cal A})$.
\item {(A2)} For all $g,h\in {\cal C}_c(X,{\cal A})$ and $y,z\in X$ holds
             \begin {equation}
             \tau_y (g(z)\tau_z h)
              = (\tau_y g)(yz)\tau_{yz}h
             \label {taucocycle}
             \end {equation}
\item {(A3)} Given $f\in {\cal C}_c(X,{\cal A})$, the function
             $g$ defined by $g(x)=(\tau_x f)(e)^*$ belongs to
             ${\cal C}_c(X,{\cal A})$ 
             and satisfies $(\tau_x g)(y)=(\tau_y f)(x)^*$ for all
             $x,y\in X$.
\item {(A4)} For all $x,y\in X$ and $f\in {\cal C}_c(X,{\cal A})$ is
             $||(\tau_x f)(y)||=||f(x^{-1}y)||$.
\end{description}
\end{definition}

\noindent

The following characterization of $C^*$-multipliers is proved.

\begin {theorem}
\label {xitau}
Let $X$ be a locally compact Hausdorff group and $\cal A$ a
$C^*$-algebra. There is a one-to-one correspondence between twisting
pairs $(\xi,\sigma)$, consisting of an
$\cal A$-multiplier $\xi$ of $X$ and an associated
twisted representation $\sigma$,
and projective actions $\tau$ of $X$ in ${\cal C}_c(X,{\cal A})$.
\end {theorem}

\noindent
Recall that a group action $\tau$ of $X$ as automorphisms of $\cal A$ can be
used to construct the crossed product algebra ${\cal A}\times_\tau X$.
This is done by defining a multiplication and an involution on the
linear space ${\cal C}_c(X,{\cal A})$ to make it into an involutive
algebra. The latter is then completed by closure in a $C^*$-norm.
This construction is given here in terms of projective actions
using continuous functions instead of measurable functions.
The projective action $\tau$ is shown to leave the $C^*$-norm
invariant.

\begin{example}
Let $X=\{0,1\}$ and let ${\cal A}={\bf C}$.
Then ${\cal C}_c(X,{\cal A})$ coincides with ${\bf C}^2$.
A projective action $\tau$ of $X$ as linear transformations
of ${\bf C}^2$ is defined by
\begin {equation}
\tau_0\left(\matrix {a\cr b\cr}\right)=
\left(\matrix {a\cr b\cr}\right)
\qquad\hbox{ and }
\tau_1\left(\matrix {a\cr b\cr}\right)=
\left(\matrix {e^{i\alpha}b\cr a\cr}\right)
\end {equation}
for all $a,b\in {\bf C}$, with $\alpha\in {\bf R}$ fixed.
It is easy to verify that $\tau$ is indeed a projective action.
The multiplier $\xi$ is given by $\xi(1,1)=e^{i\alpha}$ and
$\xi(x,y)=1$ otherwise.
The crossed product algebra ${\cal A}\times_\tau X$ equals
${\bf C}^2$ with product law
\begin{equation}
\left(\matrix {a\cr b\cr}\right)\times\left(\matrix {c\cr d\cr}\right)
=\left(\matrix {ac+bde^{i\alpha}\cr ad+bc\cr}\right)
\end{equation}
and involution
\begin{equation}
\left(\matrix {a\cr b\cr}\right)^*
=\left(\matrix {\overline a\cr \overline b e^{-i\alpha}\cr}\right)
\end{equation}
It is an abelian algebra with
unit $\displaystyle \left(\matrix {1\cr 0\cr}\right)$.
\qed
\end{example}

Recall that a representation $\pi$ of $\cal A$ in a Hilbert space $\cal H$ is 
$X$-covariant if there exists a unitary representation $x\rightarrow U(x)$ of 
$X$ such that $\pi(\sigma_xa)=U(x)\pi(a)U(x)^*$ holds for all $x\in X$ and
$a\in {\cal A}$. The crossed product algebra ${\cal A}\times_\tau X$ has the
basic property that there is a relation between its *-representations and the 
covariant *-representations of $\cal A$. This property is preserved in the 
following form.

\begin {theorem}
\label {weylop}
Let $X$ be a locally compact Hausdorff group. Let $\cal A$
be a $C^*$-algebra. Let $\xi$ be an $\cal A$-multiplier of $X$,
$\sigma$ a twisted representation of $X$
associated with $\xi$, and let $\tau$ be the
projective action determined by $\xi$ and $\sigma$
via (\ref {taudef}).
${\cal A}$ can be identified with a
sub-$C^*$-algebra of the multiplier algebra $M({\cal A}\times_\tau X)$.
There exists a map $W$ of $X$ into the unitary elements of
$M({\cal A}\times_\tau X)$ satisfying
\begin {equation}
\tau_xf=W(x)f,
\qquad x\in X, f\in {\cal A}\times_\tau X
\end {equation}
and
\begin {equation}
\sigma_xa=W(x)aW(x)^*,
\qquad x\in X,a\in {\cal A}
\label {sigmaW}
\end {equation}
and
\begin {equation}
W(x)W(y)=\xi(x,y)W(xy),
\qquad x,y\in X
\label {wlaw}
\end {equation}
\end {theorem}

\noindent
The $W(x)$ generalize the Weyl operators,
see e.g.~\cite {PD90} and the discussion below.

\begin{example}
Let $X=\{1,i,j,k\}$ be the Vierergruppe von Klein. One has
$i^2=j^2=k^2=1$ and $ij=ji=k$. A cocycle $\xi$ with values in $\bf C$
is defined
by $\xi(i,j)=\xi(j,k)=\xi(k,i)=i$, $\xi(j,i)=\xi(k,j)=\xi(i,k)=-i$,
and $\xi=1$ otherwise (note that the symbol $i$ is used with 2 different meanings).
Take $\cal A={\bf C}$.
The crossed product algebra ${\cal A}\times_\tau X$ is the
algebra of complex 4-by-4 matrices of the form
\begin{equation}
\left(\matrix{
a &b   &c   &d\cr
b &a   &-id &ic\cr
c &id  &a   &-ib\cr
d &-ic &ib  &a\cr
}\right)
=aW(1)+bW(i)+cW(j)+dW(k)
\end{equation}
The generalized Weyl operators satisfy the same
commutation relations as the Pauli matrices.
In particular, ${\cal A}\times_\tau X$ is
isomorphic to the algebra of all complex 2-by-2 matrices.
\qed
\end{example}

Assume now that a *-representation $\pi$ of ${\cal A}\times_\tau X$
in a Hilbert space $\cal H$ with cyclic vector $\Omega$ is given.
Then, by the previous theorem, there exists a map $x\rightarrow U(x)$
of $X$ into the unitary operators of $\cal H$
such that
\begin {equation}
\pi(\tau_xf)=U(x)\pi(f),
\qquad x\in X,f\in {\cal A}\times_\tau X
\label {uxleft}
\end {equation}
and
\begin {equation}
\pi(\sigma_xa)=U(x)\pi(a)U(x)^*,
\qquad x\in X,a\in {\cal A}
\label {covres}
\end {equation}
and
\begin {equation}
U(x)U(y)=\pi(\xi(x,y))U(xy),
\qquad x,y\in X
\label {ulaw}
\end {equation}
Indeed, it suffices to take $U(x)=\pi(W(x))$ for all $x\in X$.
This shows that any *-representation of ${\cal A}\times_\tau X$
with cyclic vector
defines a (generalized) covariant representation of $\cal A$.
The obvious reason to apply the previous theorem in quantum mechanics is
for the  construction of covariant representations. It can also be used
to construct representations of the CCR, as is shown below.

Recall that a symplectic space $H,s$ is a real vector space
$H$ equipped with a symplectic form $s$,
i.e., a real bilinear antisymmetric form which is non-degenerate:
$s(x,y)=0$ for all $y\in H$ implies $x=0$. A quantization of
$H$ is a map
$x\in H\rightarrow W(x)$ of $H$ into a $C^*$-algebra
satisfying the following CCR.
\begin {equation}
W(x)W(y)=e^{is(x,y)}W(x+y),
\qquad x,y\in H
\end {equation}
The unique $C^*$-algebra generated by the unitary elements
$W(x)$, $x\in X$, is called the $C^*$-algebra of CCR.
The vector space $H$ can be considered as an abelian group
for the addition. It is locally compact for the discrete topology.
A multiplier $\xi$ of $H$ with values in $\bf C$ is defined by
\begin {equation}
\xi(x,y)=e^{is(x,y)}
\end {equation}
The associated twisted representation $\sigma$ is trivial.
This shows that (\ref {wlaw}) is a generalization of the standard CCR.

A simple but nontrivial example of this quantization is the quantum
spacetime of \cite {DFR94}, \cite {DFR95}. This example is worked out in
the last section of the paper.

%%%%%%%%%%%%%%%%%%%%%%%%%%%%%%%%%%%%%%%%%%%%%%%%%%%%%%%%%%%%%%%%%%%%%%%%%%%
\section {Proof of Theorem \ref {xitau}}

First assume that $\xi$ and $\sigma$ are given, and let $\tau$
be defined by (\ref {taudef}). The continuity (M4) is needed
to show that $\tau_x$ maps ${\cal C}_c(X,{\cal A})$ into itself.
The properties (A1) to (A4) can be verified by straightforward
calculation. Note that, in order to prove that the function $g$
of (A3) is continuous one needs again (M4).

The remainder of this section deals with the proof of theorem \ref
{xitau} in the other direction. It is assumed in subsequent subsections
that a map $\tau$ of $X$ into the linear transformations of
${\cal C}_c(X,{\cal A})$, satisfying the axioms (A1) to (A4), is given.
The study of its properties will lead to the proof that $\tau$ is a
projective action associated with a $C^*$-multiplier $\xi$.

%%%%%%%%%%%%%%%%%%%%%%%%%%%%%%%%%%%%%%%%%%%%%%%%%%%%%%%%%%%%%%%%%%%%%%%%%%%
\subsection {Projective actions}

\begin {notation}
Given $\lambda\in {\cal C}_c(X)$ and $a\in {\cal A}$,
let $W^a(\lambda)$ denote the function given by
\begin {equation}
W^a(\lambda)(x)=\lambda(x)a
\end {equation}
\end {notation}
Clearly, $W^a(\lambda)$ belongs to ${\cal C}_c(X,{\cal A})$.

\begin {proposition}
\label {invertible}
For each $x\in X$ and $f\in {\cal C}_c(X,{\cal A})$
there exists $g\in {\cal C}_c(X,{\cal A})$ such that $f=\tau_xg$.
\end {proposition}

\begin {proof}
Take $z=y^{-1}$ in (\ref {taucocycle}). One obtains
\begin {equation}
\tau_x (h(x^{-1})\tau_{x^{-1}} f)
= (\tau_x h)(e)f
\end {equation}

Fix $a\in {\cal A}$ and
let us construct a function $h$ for which $(\tau_x h)(e)=a$
holds.
Let $\lambda\in {\cal C}_c(X)$ be such that $\lambda(x)=1$. Let
\begin {equation}
h(y)=\tau_yW^{a^*}(\lambda)(e)^*,
\qquad y\in X
\end {equation}
Then by axiom (A3) one obtains
\begin {equation}
\tau_x h(e)=W^{a^*}(\lambda)(x)^*=\overline {\lambda(x)}a=a
\end {equation}

This shows that for each $a\in {\cal A}$ there exists a function
$g\in {\cal C}_c(X,{\cal A})$ such that $af=\tau_xg$. Now
let $(u_\alpha)_\alpha$ be an approximate unit of $\cal A$
and let $g_\alpha$ be such that $u_\alpha f=\tau_xg_\alpha$.
From the following estimate, obtained using (A4),
\begin {eqnarray}
||g_\alpha(y)-g_\beta(y)||
&=&||\tau_xg_\alpha(xy)-\tau_xg_\beta(xy)||\cr
&=&||(u_\alpha -u_\beta) f(xy)||
\label {crprop3}
\end {eqnarray}
follows that the elements $(g_\alpha(y))_\alpha$
converge to some function $g$. The support $S(g)$
of $g$ equals $x^{-1}S(f)$. The function $g$ is continuous
as is obvious from
\begin {eqnarray}
||g(y)-g(z)||
&=&\lim_\alpha||g_\alpha(y)-g_\alpha(z)||\cr
&=&\lim_\alpha||\tau_xg_\alpha(xy)-\tau_xg_\alpha(xz)||\cr
&=&\lim_\alpha||u_\alpha(f(xy)-f(xz))||\cr
&=& ||f(xy)-f(xz)||
\end {eqnarray}
Hence $g$ belongs to ${\cal C}_c(X,{\cal A})$.
Finally, from
\begin {equation}
\lim_\alpha ||\tau_xg_\alpha(y)-\tau_xg(y)||
=\lim_\alpha ||g_\alpha(x^{-1}y)-g(x^{-1}y)||=0
\end {equation}
follows that $f=\tau_xg$. This ends the proof.
\qed
\end {proof}

\begin {corollary}
\label {cancel}
             If for a given $x\in X$ and $a,b\in {\cal A}$ one has
             $a\tau_xf(e)=b\tau_xf(e)$ for all $f\in {\cal C}_c(X,{\cal A})$
             then $a=b$ follows.
\end {corollary}

\begin {proof}
As a consequence of the previous proposition one may assume that $x=e$.
Now take $f=W^c(\lambda)$ with $c\in {\cal A}$ and $\lambda\in {\cal C}_c(X)$.
There follows $\lambda(e)ac=\lambda(e)bc$. Since $c$ and
$\lambda$ are arbitrary there follows $a=b$.
\qed
\end {proof}

%%%%%%%%%%%%%%%%%%%%%%%%%%%%%%%%%%%%%%%%%%%%%%%%%%%%%%%%%%%%%%%%%%%%%%%%%%%
\subsection {Existence of the $C^*$-multiplier}

The properties of $\tau$ proven in the previous subsection can be used
to establish the existence of a $C^*$-multiplier $\xi$. The proof that
$\xi$ satisfies all axioms defining a $C^*$-multiplier is completed
later on.

\begin {proposition}
\label {xiprop}
There exists a map $\xi:\ X\times X\rightarrow M({\cal A})$
such that for all $f\in {\cal C}_c(X,{\cal A})$ holds
\begin {equation}
\tau_y\tau_zf=\xi(y,z)\tau_{yz}f,
\qquad y,z\in X
\label {tautau}
\end {equation}
\end {proposition}

\begin {proof}
For a fixed $z$ one can select a $\lambda\in {\cal C}_c(X)$ for which
$\lambda(z)\not=0$. Let $(u_\alpha)_\alpha$ be an approximate
unit of $\cal A$, and define $\xi(y,z)$ by
\begin {equation}
\xi(y,z)a=\lambda(z)^{-1}\lim_\alpha(\tau_yW^{u_\alpha}(\lambda))(yz)a,
\qquad a\in {\cal A}
\label {xidef}
\end {equation}
That the limit converges follows from the following estimates.
Let $h\in {\cal C}_c(X,{\cal A})$ be such that $\tau_{yz}h(e)=a$
(see the proof of proposition \ref {invertible}). Then, using (A2)
and (A4),
one obtains
\begin {eqnarray}
||\tau_y\left(W^{u_\alpha}(\lambda)-W^{u_\beta}(\lambda)\right)(yz)a||
&=&||\tau_y\left(W^{u_\alpha}(\lambda)-W^{u_\beta}(\lambda)\right)(yz)
\tau_{yz}h(e)||\cr
&=&||\tau_y\left(\left(W^{u_\alpha}(\lambda)-W^{u_\beta}(\lambda)\right)
(z)\tau_{z}h\right)(e)||\cr
&=&||\left(W^{u_\alpha}(\lambda)-W^{u_\beta}(\lambda)\right)
(z)\tau_{z}h(y^{-1})||\cr
&=&|\lambda(z)|\,||(u_\alpha-u_\beta)\tau_{z}h(y^{-1})||
\label {convestim}
\end {eqnarray}
Hence $(\tau_yW^{u_\alpha}(\lambda)(yz)a)_\alpha$ forms a Cauchy net
converging to $\lambda(z)\xi(y,z)a$.

It is straightforward to prove that $\xi(y,z)$ belongs to $M({\cal A})$.

From (A2) one obtains for arbitrary $f\in{\cal C}_c(X,{\cal A})$
\begin {eqnarray}
(\tau_yW^{u_\alpha}(\lambda))(yz)\tau_{yz} f
&=&\tau_y(W^{u_\alpha}(\lambda)(z)\tau_zf)\cr
&=&\tau_y(\lambda(z)u_\alpha\tau_zf)\cr
&=&\lambda(z)\tau_y(u_\alpha\tau_zf)
\label {wtau}
\end {eqnarray}
From (A4) follows that $\tau_y(u_\alpha\tau_zf)(x)$
converges to $(\tau_y\tau_zf)(x)$.
Hence (\ref {wtau}) implies (\ref {tautau}).

Note that $\xi(y,z)$ does not depend on the choice
of $\lambda$. Indeed, if $\lambda(z)\not=0$ and $\mu(z)\not=0$ then
\begin {equation}
\left[\lambda(z)^{-1}(\tau_yW^{u_\alpha}(\lambda))(yz)
-\mu(z)^{-1}(\tau_yW^{u_\alpha}(\mu))(yz)\right]\tau_{yz}f=0
\end {equation}
for all $f$, because of (\ref {wtau}). By corollary \ref {cancel}
this implies
\begin {equation}
\lambda(z)^{-1}(\tau_yW^{u_\alpha}(\lambda))(yz)
-\mu(z)^{-1}(\tau_yW^{u_\alpha}(\mu))(yz)=0
\end {equation}
This shows that both candidates for the r.h.s.~of (\ref {xidef}) coincide,
i.e.~the definition of $\xi$ does not depend on the choice of $\lambda$.
\qed
\end {proof}

\begin {proposition}
\label {xiprop2}
The map $\xi$ of the previous proposition satisfies 
\begin {equation}
\xi(y,z)^*\xi(y,z)=1,
\qquad
y,z\in X
\label {xiiso}
\end {equation}
\end {proposition}

\begin {proof}
From (A4) and (\ref {tautau}) follows
\begin {eqnarray}
||\xi(y,z)\tau_{yz}f(x)||
&=&||\tau_y\tau_zf(x)||\cr
&=&||f(z^{-1}y^{-1}x)||\cr
&=&||\tau_{yz}f(x)||
\label {tempprop2}
\end {eqnarray}
Take $a\in {\cal A}$
and $\lambda\in {\cal C}_c(X)$ arbitrarily, and let $f$ be defined by
\begin {equation}
f(x)=\tau_xW^{a^*}(\lambda)(e)^*
\end{equation}
Then (A3) implies that
\begin {equation}
\tau_{yz}f(e)=W^{a^*}(\lambda)(yz)^*=
\overline {\lambda(yz)}a
\end {equation}
Equation (\ref {tempprop2}) becomes
\begin {equation}
|\lambda(yz)|\,||\xi(y,z)a||=|\lambda(yz)|\,||a||
\end {equation}
Since $\lambda$ is arbitrary one concludes that
$\xi(y,z)$ is an isometry of $\cal A$.
\qed
\end {proof}

Note that the map $\xi$ satisfies
\begin {equation}
\xi(e,y)=\xi(x,e)={\bf 1},
\qquad x,y\in X
\label {xiunit}
\end {equation}
This is an immediate consequence of (\ref {tautau}) and corollary
\ref {cancel}.

%%%%%%%%%%%%%%%%%%%%%%%%%%%%%%%%%%%%%%%%%%%%%%%%%%%%%%%%%%%%%%%%%%%%%%%%%%%
\subsection {Existence of the associated twisted representation}

One can identify ${\cal A}$
with a class of linear transformations
of ${\cal C}_c(X,{\cal A})$ in the following way.

\begin {notation}
For each $a\in {\cal A}$ introduce a linear transformation
$\zeta^a$ of ${\cal C}_c(X,{\cal A})$ by
\begin {equation}
\zeta^af(x)=af(x),
\qquad x\in X,f\in {\cal C}_c(X,{\cal A})
\end {equation}
\end  {notation}

\noindent
These linear transformations will later on become elements of the
multiplier algebra of the crossed product algebra
${\cal A}\times_\tau X$.

Let us now show that there exists a deformed representation $\sigma$
of $X$ in $\cal A$.

\begin {proposition}
\label {sigmaexists}
There exists a map $\sigma$ of $X$ into the *-homomorphisms
of $\cal A$ such that $\sigma_e=1$,
and
\begin {equation}
\sigma_xf(x^{-1}y)=\tau_xf(y)\xi(x,x^{-1}y)^*,
\qquad x,y\in X, f\in {\cal C}_c(X,{\cal A})
\label {almosttaudef}
\end {equation}
and
\begin {equation}
\tau_x\circ\zeta^a=\zeta^{\sigma_xa}\circ\tau_x,
\qquad x\in X
\label {zetacon}
\end {equation}
For each $a\in {\cal A}$ the map $x\in X\rightarrow\sigma_xa$
is continuous.
\end {proposition}

\begin {proof}
Let $\lambda\in {\cal C}_c(X)$ be such that $\lambda(e)\not=0$.
Define $\sigma_x$ by 
\begin {equation}
\sigma_xa={1\over\lambda(e)}\tau_xW^a(\lambda)(x),
\qquad x\in X,a\in {\cal A}
\label {sigmadeftau}
\end {equation}
Note that the definition of $\sigma_x$ does not depend on
the choice of $\lambda$. Indeed, using (A2), one obtains
for any $f\in {\cal C}_c(X,{\cal A})$
\begin {eqnarray}
{1\over\lambda(e)}\tau_xW^a(\lambda)(x)\tau_xf
&=&{1\over\lambda(e)}\tau_x(W^a(\lambda)(e)f)\cr
&=&\tau_x(af)
\label {tauzetaint}
\end {eqnarray}
By corollary \ref {cancel} this implies
that the definition of $\sigma_x a$ does
not depend on the choice of $\lambda$.
Equation (\ref {tauzetaint}) shows at once that (\ref {zetacon}) holds.

Obviously, one has
\begin {equation}
\sigma_ea={1\over\lambda(e)}\tau_eW^a(\lambda)(e)=a
\end {equation}
This proves that $\sigma_e$ is the identity transformation of $\cal A$.

Now take any $f\in {\cal C}_c(X,{\cal A})$ and calculate
\begin {eqnarray}
\sigma_xf(x^{-1}y)
&=&{1\over\lambda(e)}\tau_xW^{f(x^{-1}y)}(\lambda)(x)\cr
&=&{1\over\lambda(e)}\tau_x(f(x^{-1}y)\tau_{x^{-1}y}h)(x)
\end {eqnarray}
with $h$ such that $W^e(\lambda)=\tau_{x^{-1}y}h$
(this exists because of proposition \ref {invertible}).
Using (A2) there follows
\begin {eqnarray}
\sigma_xf(x^{-1}y)
&=&{1\over\lambda(e)}\tau_xf(y)\tau_yh(x)\cr
&=&{1\over\lambda(e)}\tau_xf(y)\xi(x,x^{-1}y)^*\tau_xW^e(\lambda)(x)\cr
&=&\tau_xf(y)\xi(x,x^{-1}y)^*
\end {eqnarray}
This is (\ref {almosttaudef}).

Because of (A3) the map $x\rightarrow\tau_xf(y)$ is continuous
for all $y\in X$ and $f\in {\cal C}_c(X,{\cal A})$. Hence also the map
\begin {equation}
x\rightarrow \sigma_xa={1\over\lambda(e)}\tau_xW^a(\lambda)(x)
\end {equation}
 is continuous.

Finally, let us show that $\sigma_x$ is a *-homomorphism of $\cal A$.
Clearly, $\sigma_x$ is a linear transformation.

Given $a,b\in {\cal A}$ , using (A2), one finds
\begin {eqnarray}
\sigma_x(ab)
&=&{1\over\lambda(e)}\tau_x W^{ab}(\lambda)(x)\cr
&=&{1\over\lambda(e)^2}\tau_x(W^a(\lambda)(e)W^b(\lambda))(x)\cr
&=&{1\over\lambda(e)^2}\tau_xW^a(\lambda)(x)\tau_xW^b(\lambda)(x)\cr
&=&(\sigma_xa)(\sigma_xb)
\end {eqnarray}
This shows that $\sigma_x$ is a homomorphism.

Let $g(x)=\tau_xW^{a^*}(\lambda)(e)^*$.
Then by (A3) one has
\begin {equation}
\sigma_xa^*={1\over\lambda(e)}\tau_xW^{a^*}(\lambda)(x)
={1\over\lambda(e)}\tau_xg(x)^*
\label {sigstar}
\end {equation}
For any $h\in {\cal C}_c(X,{\cal A})$ holds, using (A2) twice,
\begin {eqnarray}
\tau_xg(x)\tau_xh
&=&\tau_x(g(e)h)\cr
&=&\tau_x(W^a(\overline\lambda)(e)h)\cr
&=&\tau_x W^a(\overline\lambda)(x)\tau_xh
\end {eqnarray}
By corollary \ref {cancel} there follows that
$\tau_xg(x)=\tau_x W^a(\overline\lambda)(x)$.
Hence (\ref {sigstar}) becomes
\begin {eqnarray}
(\sigma_xa^*)^*
&=&{1\over\overline {\lambda(e)}}\tau_x W^a(\overline\lambda)(x)\cr
&=&\sigma_xa
\end {eqnarray}
This ends the proof of the proposition.
\qed
\end {proof}

\begin {corollary}
\label {xiunifcont}
For each $a\in {\cal A}$ and $x\in X$ the map $y\rightarrow a\xi(x,y)$
is continuous.
\end {corollary}

\begin {proof}
Fix $\lambda\in{\cal C}_c(X)$.
Let $b=\sigma_x^{-1}a$ ($\sigma_x$ is invertible --- see corollary \ref{sigmainvert}
below). Let $f=W^b(\lambda)$. Then (\ref {almosttaudef}) implies
\begin {equation}
\tau_xW^b(\lambda)(xy)=\lambda(y)a\xi(x,y)
\end {equation}
The function $y\rightarrow \tau_xW^b(\lambda)(xy)$ is continuous.
Hence $y\rightarrow a\xi(x,y)$ is continuous on the set
of $y$ for which $\lambda(y)\not=0$. Since $\lambda$ is arbitrary
overall continuity follows.

\qed
\end {proof}

\begin {corollary}
The map $x\rightarrow a\xi(x,x^{-1})$ is continuous for all
$a\in {\cal A}$.
\end {corollary}

\begin {proof}
Let $\lambda$ and $b$ be as in the previous lemma.
One obtains
\begin{equation}
\tau_xW^b(\lambda)(e)=\lambda(x^{-1})a\xi(x,x^{-1})
\end{equation}
The l.h.s.~of this expression is a continuous function
of $x$ because of (A3). Hence $x\rightarrow a\xi(x,x^{-1})$
is continuous on the set of $x$ for which $\lambda(x^{-1})\not=0$.
Since $\lambda$ is arbitrary
overall continuity follows.
\qed
\end {proof}
 
\begin {proposition}
$\sigma$ satisfies (\ref {sigmadef}).
\end {proposition}

\begin {proof}
Let $a\in {\cal A}$.
From the definition (\ref {sigmadeftau}) follows
\begin {equation}
\sigma_x(\xi(y,z)a)={1\over\lambda(e)}\tau_xW^{\xi(y,z)a}(\lambda)(x)
\label {sigmatemp}
\end {equation}
for any $\lambda$ for which $\lambda(e)\not=0$. We may assume as
well that $\lambda(z)\not=0$. Then, by (\ref {xidef}),
\begin {equation}
W^{\xi(y,z)a}(\lambda)={1\over\lambda(z)}\lim_\alpha
\tau_yW^{u_\alpha}(\lambda)(yz)W^a(\lambda)
\end {equation}
Hence (\ref {sigmatemp}) becomes
\begin {equation}
\sigma_x(\xi(y,z)a)=
{1\over\lambda(e)}{1\over\lambda(z)}\lim_\alpha
\tau_x(\tau_yW^{u_\alpha}(\lambda)(yz)W^a(\lambda))(x)
\end {equation}

By proposition \ref {invertible} there exists a function $h$
in ${\cal C}_c(X,{\cal A})$ such that $W^a(\lambda)=\tau_{yz}h$.
Using this together with (A2) gives
\begin {eqnarray}
\sigma_x(\xi(y,z)a)
&=&{1\over\lambda(e)}{1\over \lambda(z)}\lim_\alpha
\tau_x\left(\tau_yW^{u_\alpha}(\lambda)(yz)\tau_{yz}h\right)(x)\cr
&=&{1\over\lambda(e)}{1\over \lambda(z)}\lim_\alpha
\tau_x\tau_yW^{u_\alpha}(\lambda)(xyz)
\tau_{xyz}h(x)\cr
&=&{1\over\lambda(e)}{1\over \lambda(z)}\lim_\alpha
\xi(x,y)\tau_{xy}W^{u_\alpha}(\lambda)(xyz)
\xi(x,yz)^*\tau_xW^a(\lambda)(x)\cr
&=&\xi(x,y)\xi(xy,z)\xi(x,yz)^*\sigma_xa
\end {eqnarray}
Because $a$ is arbitrary, (\ref {sigmadef}) follows.
\qed
\end {proof}

\begin {corollary}
For all $x,y\in X$ is
\begin {equation}
\xi(x,y)\xi(x,y)^*=1
\label {unitxi}
\end {equation}
\end {corollary}

\begin {proof}
Take $z=e$ in (\ref {sigmadef}). Then (\ref {unitxi}) follows.
\qed
\end {proof}

\begin {proposition}
$\sigma$ satisfies (\ref {projtrans}).
\end {proposition}

\begin {proof}
One calculates
\begin {eqnarray}
\sigma_x\sigma_ya
&=&{1\over\lambda(e)}\tau_xg^{\sigma_ya}(x)\cr
&=&{1\over\lambda(e)}\tau_x
(W^e(\lambda)\sigma_ya)(x)\cr
&=&{1\over\lambda(e)}\tau_x
\left(W^e(\lambda){1\over\lambda(e)}\tau_yW^a(\lambda)(y)\right)(x)
\end {eqnarray}
From proposition \ref {invertible} follows that
$W^e(\lambda)=\tau_yh$ for some $h\in {\cal C}_c(X,{\cal A})$.
Hence the previous expression becomes
\begin {eqnarray}
\sigma_x\sigma_ya
&=&{1\over\lambda(e)}\tau_x
\left({1\over\lambda(e)}\tau_yW^a(\lambda)(y)\tau_yh\right)(x)
\end {eqnarray}

Now apply (A2) to obtain
\begin {eqnarray}
\sigma_x\sigma_ya
&=&{1\over\lambda(e)^2}\tau_x\tau_yW^a(\lambda)(xy)
\tau_{xy}h(x)\cr
&=&\xi(x,y)(\sigma_{xy}a)\xi(x,y)^*
\end {eqnarray}
To obtain the latter, use is made of
\begin {eqnarray}
{1\over\lambda(e)}\tau_{xy}h(x)
&=&{1\over\lambda(e)}\xi(x,y)^*\tau_x\tau_yh(x)\cr
&=&{1\over\lambda(e)}\xi(x,y)^*\tau_xW^e(\lambda)(x)\cr
&=&\xi(x,y)^*
\end {eqnarray}
\qed
\end {proof}

\begin {corollary}
\label{sigmainvert}
$\sigma_x$ is an automorphism of $\cal A$ for any $x\in X$.
\end {corollary}

\begin {proof}
Proposition \ref {sigmaexists} shows that $\sigma_x$ is a *-homomorphism.
Invertibility follows from the previous proposition by taking $x=y^{-1}$.
One obtains
\begin {equation}
(\sigma_y)^{-1}a=\xi(y^{-1},y)^*(\sigma_{y^{-1}}a)\xi(y^{-1},y),
\qquad y\in X,a\in {\cal A}
\end {equation}
\qed
\end {proof}

%%%%%%%%%%%%%%%%%%%%%%%%%%%%%%%%%%%%%%%%%%%%%%%%%%%%%%%%%%%%%%%%%%%%%%%%%%%
\section {Construction}
\label {constr}

Let be given a map $\tau$ of $X$ into the linear
transformations of ${\cal C}_c(X,{\cal A})$.
Assume $\tau$ satisfies the axioms (A1) to (A4).

%%%%%%%%%%%%%%%%%%%%%%%%%%%%%%%%%%%%%%%%%%%%%%%%%%%%%%%%%%%%%%%%%%%%%%%%%%%
\subsection {Product and involution}

Consider ${\cal C}_c(X,{\cal A})$ as a linear space.
In what follows a product law and an involution are defined
which make ${\cal C}_c(X,{\cal A})$ into a *-algebra.
In the next subsection this algebra is completed in norm. In the final
subsections an approximate unit and the enveloping $C^*$-algebra
are constructed.

Define a product on ${\cal C}_c(X,{\cal A})$ by
\begin {equation}
(fg)(x)=\int_X\hbox{ d}y\ f(y)(\tau_y g)(x)
\end {equation}

\begin {lemma}
If $f$ and $g$ belong to ${\cal C}_c(X,{\cal A})$
then also $fg$ belongs to ${\cal C}_c(X,{\cal A})$.
\end {lemma}

\begin {proof}
From assumption (A4) follows that the
support of $\tau_yg$ satisfies $S(\tau_yg)=yS(g)$.
Hence
\begin {equation}
S(fg)\subset\cup_{y\in S(f)}S(\tau_yg)
=\cup_{y\in S(f)}\,yS(g)
\end{equation}
From the continuity of multiplication and inverse
and the fact that $S(f)$ and $S(g)$ are compact
follows that also
$\displaystyle
\cup_{y\in S(f)}S(g)
$
is compact. Hence $fg$ has a compact support.
Continuity of $fg$ is obvious.
\qed
\end {proof}

\begin {proposition}
\label {leftmullem}
\begin {equation}
\tau_x(fg)=(\tau_x f)g,
\qquad f,g\in {\cal C}_c(X,{\cal A}), x\in X
\label {prodrule}
\end {equation}
\end {proposition}

\begin {proof}
Observe that, using (A2),
\begin {eqnarray}
(\tau_x(fg))(y)
&=&\tau_x\left(\int_X\hbox{ d}z\ f(z)\tau_zg\right)(y)\cr
&=&\int_X\hbox{ d}z\ \tau_x\left(f(z)\tau_zg\right)(y)\cr
&=&\int_X\hbox{ d}z\ (\tau_x f)(z)(\tau_{z}g)(y)\cr
&=&((\tau_x f)g)(y)
\label {autrule}
\end{eqnarray}
Hence one concludes (\ref {prodrule}).
\qed
\end {proof}

It is now easy to show that the product is associative.
Indeed, one has
\begin {eqnarray}
((fg)h)(x)
&=&\int_X\hbox{ d}y\ (fg)(y)(\tau_y h)(x)\cr
&=&\int_X\hbox{ d}y\int_X\hbox{ d}z
\ f(z)(\tau_z g)(y)(\tau_y h)(x)\cr
&=&\int_X\hbox{ d}z
\ f(z)((\tau_z g)h)(x)\cr
&=&\int_X\hbox{ d}z
\ f(z)(\tau_z (gh))(x)\cr
&=&(f(gh))(x)
\end {eqnarray}

Next define an involution of ${\cal C}_c(X,{\cal A})$.
The adjoint $f^*$ of a function $f\in {\cal C}_c(X,{\cal A})$
is defined by
\begin {equation}
f^*(x)=\Delta(x)^{-1}\tau_x f(e)^*,
\qquad x\in X
\end {equation}
(the modular function of $X$ is denoted $\Delta$).
$f^*$ belongs again to ${\cal C}_c(X,{\cal A})$
by assumption (A3).
One has
\begin {equation}
f^{**}(x)=\Delta(x)^{-1}\tau_x\Delta^{-1}g(e)^*
\end {equation}
with $g(x)=\tau_x f(e)^*$.
Using (A2) one proves that
\begin {equation}
(\tau_x\Delta^{-1}g)(y)=\Delta^{-1}(x^{-1}y)\tau_xg(y)
\end {equation}
Hence $f^{**}(x)=\tau_xg(e)^*$ follows.
Using (A3) this implies $f^{**}(x)=f(x)$.

Obviously, the involution is compatible with the linear structure of
${\cal C}_c(X,{\cal A})$.
It is also compatible with the product. Indeed, using (A2) and (A3),
\begin {eqnarray}
(fg)^*(x)
&=&\Delta(x)^{-1}\tau_x (fg)(e)^*\cr
&=&\Delta(x)^{-1}\bigg(\tau_x 
\int_X\hbox{ d}y\ f(y)\tau_y g
\bigg)(e)^*\cr
&=&\Delta(x)^{-1}\int_X\hbox{ d}y\ \big(\tau_x \big(
f(y)\tau_y g
\big)\big)(e)^*\cr
&=&\Delta(x)^{-1}\int_X\hbox{ d}y\ \big(
(\tau_x f)(y)(\tau_{y} g)(e)
\big)^*\cr
&=&\Delta(x)^{-1}\int_X\hbox{ d}y
\ (\tau_{y} g)(e)^*
(\tau_x f)(y)^*\cr
&=&\Delta(x)^{-1}\int_X\hbox{ d}y\ \Delta(y)g^*(y)(\tau_x f)(y)^*\cr
&=&\int_X\hbox{ d}y\ g^*(y)\Delta(x^{-1}y)(\tau_y h)(x)
\end {eqnarray}
with $h(x)=(\tau_x f)(e)^*=\Delta(x)f^*(x)$.
There follows
\begin {eqnarray}
(fg)^*(x)
&=&\int_X\hbox{ d}y\ g^*(y)\Delta(x^{-1}y)(\tau_y\Delta f^*)(x)\cr
&=&(g^*f^*)(x)
\end {eqnarray}
This shows the compatibility of the involution and the multiplication.

%%%%%%%%%%%%%%%%%%%%%%%%%%%%%%%%%%%%%%%%%%%%%%%%%%%%%%%%%%%%%%%%%%%%%%%%%%%
\subsection {Norm completion}

A norm is defined on ${\cal C}_c(X,{\cal A})$ by
\begin {equation}
||f||_1=\int_X\hbox{ d}x\ ||f(x)||
\end {equation}

Let us show that this norm is compatible with the product
and the involution of ${\cal C}_c(X,{\cal A})$.

One verifies, using (A4), that
\begin {eqnarray}
||fg||_1&=&\int_X\hbox{ d}x\ ||(fg)(x)||\cr
&=&\int_X\hbox{ d}x\ ||\int_X\hbox{ d}y\ f(y)(\tau_y g)(x)||\cr
&\le&\int_X\hbox{ d}x\ \int_X\hbox{ d}y\ ||f(y)||\,||(\tau_y g)(x)||\cr
&=&\int_X\hbox{ d}x\ \int_X\hbox{ d}y\ ||f(y)||\,||g(y^{-1}x)||\cr
&=&\int_X\hbox{ d}x\ \int_X\hbox{ d}y\ ||f(y)||\,||g(x)||\cr
&=&||f||_1\,||g||_1
\end {eqnarray}
Hence the product is continuous for this norm.
Again using (A4), one calculates
\begin {eqnarray}
||f^*||_1
&=&\int_X\hbox{ d}x \ ||f^*(x)||\cr
&=&\int_X\hbox{ d}x \ \Delta(x)^{-1} \, ||\tau_x f(e)||\cr
&=&\int_X\hbox{ d}x \ \Delta(x)^{-1} \, ||f(x^{-1})||\cr
&=&\int_X\hbox{ d}x \ ||f(x)||\cr
&=&||f||_1
\end {eqnarray}

\noindent
Note that $||\tau_xf||_1=||f||_1$,
i.e.~the projective action $\tau$ leaves the $||\cdot||_1$-norm invariant.

One concludes that the completion of ${\cal C}_c(X,{\cal A})$ in the
$||\cdot||_1$-norm is an involutive Banach algebra. It is denoted
${\cal L}_1(X,{\cal A},\tau)$, the algebra of integrable functions of $X$
with values in ${\cal A}$. In many places, the
notation ${\cal L}_1({\cal A},X)$ is used. 
The notation of \cite {BS70} would read here
${\cal L}_1({\cal A},X;\xi,\sigma)$.
It is however more natural to
see the $C^*$-algebra $\cal A$ as a generalization of the complex numbers
and to adapt a notation compatible with ${\cal L}_1(X,{\bf C})$.

%%%%%%%%%%%%%%%%%%%%%%%%%%%%%%%%%%%%%%%%%%%%%%%%%%%%%%%%%%%%%%%%%%%%%%%%%%%
\subsection {Approximate unit}

Let $(u_\alpha)_\alpha$ be an approximate unit of $\cal A$.
Select for each neighborhood $v$ of $e\in X$ a continuous
non-negative function
$\delta_v$ with support in $v$ and normalized to one:
\begin {equation}
\int_X\hbox{ d}x\,\delta_v(x)=1
\end {equation}

$\big(W^{u_\alpha}(\delta_v)\big)_{\alpha,v}$ is an approximate unit of
${\cal C}_c(X,{\cal A})$.
Indeed, one has
\begin {eqnarray}
||W^{u_\alpha}(\delta_v) g -g||_1
&=&\int_X\hbox{ d}x\ ||(W^{u_\alpha}(\delta_v) g)(x)-g(x)||\cr
&=&\int_X\hbox{ d}x\ 
||\int_X\hbox{ d}y\ \delta_v(y)u_\alpha(\tau_y g)(x)-g(x)||\cr
&\le&\int_X\hbox{ d}x\ \int_X\hbox{ d}y
\ \delta_v(y)||u_\alpha(\tau_y g)(x)-g(x)||\cr
&\le&\int_X\hbox{ d}x\ \int_X\hbox{ d}y
\ \delta_v(y)||u_\alpha[(\tau_y g)(x)-g(x)]||\cr
&+&\int_X\hbox{ d}x\ \int_X\hbox{ d}y
\ \delta_v(y)||(u_\alpha-1)g(x)||\cr
&\le&\int_X\hbox{ d}x\ \int_X\hbox{ d}y
\ \delta_v(y)||(\tau_y g)(x)-g(x)||\cr
&+&\int_X\hbox{ d}x
\ ||(u_\alpha-1)g(x)||
\end {eqnarray}
The latter expression can be made arbitrary small (note that
from (A3) follows that $y\rightarrow\tau_yg(x)$ is continuous;
use further that $\tau_yg$ is continuous with compact support).

Using (\ref {taudef}) one obtains
\begin {eqnarray}
& &||gW^{u_\alpha}(\delta_v) -g||_1\cr
&=&\int_X\hbox{ d}x\ ||(gW^{u_\alpha}(\delta_v))(x)-g(x)||\cr
&=&\int_X\hbox{ d}x\ 
||\int_X\hbox { d}y\ g(y)\tau_yW^{u_\alpha}(\delta_v)(x)
-g(x)||\cr
&=&\int_X\hbox{ d}x\ 
||\int_X\hbox { d}y\ \left(g(xy)\tau_{xy}W^{u_\alpha}(\delta_v)(x)
-g(x)\Delta^{-1}(y)\delta_v(y^{-1})\right)||\cr
&\le&\int_X\hbox{ d}x\ \int_X\hbox { d}y\ 
||(g(xy)-g(x)||\,||\tau_{xy}W^{u_\alpha}(\delta_v)(x)||\cr
& &+
\int_X\hbox{ d}x\ \int_X\hbox { d}y\ 
||g(x)\left(\tau_{xy}W^{u_\alpha}(\delta_v)(x)
-\Delta^{-1}(y)\delta_v(y^{-1})\right)||
\end {eqnarray}
Each of these two terms can be made arbitrary small.
Hence $\big(W^{u_\alpha}(\delta_v)\big)_{\alpha,v}$
is also a right approximate unit.

%%%%%%%%%%%%%%%%%%%%%%%%%%%%%%%%%%%%%%%%%%%%%%%%%%%%%%%%%%%%%%%%%%%%%%%%%%%
\subsection {The crossed product algebra}
\label {crossprod}

The crossed product of $\cal A$ with $X$ is denoted
\begin {equation}
{\cal A}\times_\tau X
\end {equation}
It is defined as the enveloping $C^*$-algebra of
${\cal L}_1(X,{\cal A},\tau)$, see \cite {DJ77}, Section 2.7.
The $C^*$-norm is given by
\begin {equation}
||f||=\sup\{\omega(f^*f)^{1/2}:
\ \omega\hbox{ positive normalized functional of }
{\cal L}_1(X,{\cal A},\tau)\}
\label {cstarnorm}
\end {equation}
The proof that $||\cdot||$ is nondegenerate goes as follows.
We first need

\begin {proposition}
Fix $f\in {\cal C}_c(X,{\cal A})$.
Any state $\omega_0$ of $\cal A$ extends to a positive
linear form $\omega_f$ of ${\cal L}_1(X,{\cal A},\tau)$
by
\begin {equation}
\omega_f(g)=\omega_0((f^*gf)(e))
\end {equation}
\end {proposition}

\begin {proof}
Linearity is obvious. Positivity follows from
\begin {eqnarray}
\omega_f(g^*g)&=&
\omega_0((f^*g^*gf)(e))\cr
&=&\int_X\hbox{ d}z\ \omega_0((gf)^*(z)\tau_z(gf)(e))\cr
&=&\int_X\hbox{ d}z\ \Delta(z)^{-1}\omega_0(\tau_z(gf)(e)^*\tau_z(gf)(e))\cr
&\ge& 0
\end {eqnarray}

It is straightforward to show that a constant $K(f)$ exists
such that $|\omega_f(g)|\le K(f)||g||_1$ holds for all
$g\in {\cal C}_c(X,{\cal A})$.
Hence, by continuity $\omega_f$ extends to a positive
linear functional on ${\cal L}_1(X,{\cal A},\tau)$.
\qed
\end {proof}

\begin {proposition}
The norm (\ref {cstarnorm}) is nondegenerate.
\end  {proposition}

\begin {proof}
Assume that $||g||=0$. Then $\omega_f(g^*g)=0$ for all $f\in {\cal
C}_c(X,{\cal A}$. By construction this implies
$\omega_0((f^*g^*gf)(e))=0$. Now take $\omega_0$ faithful. Then one
concludes that $(gf)(e)=0$ for all $f\in {\cal C}_c(X,{\cal A})$.
This means
\begin {equation}
\int_X\hbox{ d}y\ g(y)\tau_yf(e)=0
\end {equation}
$g=0$ follows using the same argument as in corollary \ref {cancel}.
Hence $||\cdot||$ is nondegenerate.
\qed
\end {proof}

Invariance of $||\cdot||$ under $\tau$, i.e.~$||\tau_xf||=||f||$ for all
$x\in X$ and $f\in {\cal A}\times_\tau X$,
follows from the following result, which makes use of the results
of section 3.

\begin {proposition}
\label{isoprop}
\begin {equation}
(\tau_xf)^*\tau_xg=f^*g,
\qquad
f,g\in {\cal C}_c(X,{\cal A})
\label {isotau}
\end {equation}
\end {proposition}

\begin {proof}
One calculates
\begin {eqnarray}
(\tau_xf)^*\tau_xg
&=&\int_X\hbox{ d}y\ (\tau_xf)^*(y)\tau_y\tau_xg\cr
&=&\int_X\hbox{ d}y\ \Delta(y)^{-1}\tau_y\tau_xf(e)^*\tau_y\tau_xg\cr
&=&\int_X\hbox{ d}y\ \Delta(y)^{-1}(\tau_{yx}f)(e)^*\xi(y,x)^*
\xi(y,x)\tau_{yx}g\cr
&=&\int_X\hbox{ d}y\ \Delta(y)^{-1}\tau_{yx}f(e)^*
\tau_{yx}g\cr
&=&\int_X\hbox{ d}y\ \Delta(x)
f^*(yx)\tau_{yx}g\cr
&=&\int_X\hbox{ d}y\ f^*(y)\tau_{y}g\cr
&=&f^*g
\end {eqnarray}
This shows (\ref {isotau}).
\qed
\end {proof}

%%%%%%%%%%%%%%%%%%%%%%%%%%%%%%%%%%%%%%%%%%%%%%%%%%%%%%%%%%%%%%%%%%%%%%%%%%%
\section {Proof of theorem \ref {weylop} }
\label {thm2sect}

The following result shows that $\cal A$ can be identified with
a sub-$C^*$-algebra of $M({\cal A}\times_\tau X)$.

\begin {proposition}
$\zeta$ extends to an injection of $\cal A$ into $M({\cal A}\times_\tau X)$.
One has $(\zeta^a)^*=\zeta^{a^*}$ for all $a\in {\cal A}$.
\end {proposition}

\begin {proof}
By continuity $\zeta$ extends to a linear transformation of
${\cal A}\times_\tau X$.
A straightforward calculation shows that
\begin {equation}
(\zeta^{a^*}f)^*g=f^*\zeta^ag,
\qquad f,g\in {\cal C}_c(X,{\cal A})
\end {equation}
Hence $\zeta^a$ belongs to the multiplier algebra of ${\cal A}\times_\tau X$.
\qed
\end {proof}

From now on $\zeta$ is omitted, i.e.~$\cal A$ is considered to be
a sub-$C^*$-algebra of $M({\cal A}\times_\tau X)$.

\begin {proposition}
The linear transformation $W(x)$ defined by
\begin {equation}
W(x)f=\tau_xf,
\qquad x\in X,f\in {\cal A}\times_\tau X
\end {equation}
belongs to $M({\cal A}\times_\tau X)$.
\end {proposition}

\begin {proof}
For all $f,g\in {\cal A}\times_\tau X$ is
\begin {eqnarray}
& &\left( \xi(x^{-1},x)^*W(x^{-1})f\right)^*g\cr
&=&\int_X\hbox{ d}z\ \left(\xi(x^{-1},x)^*W(x^{-1})f\right)^*(z)\tau_zg\cr
&=&\int_X\hbox{ d}z\ \Delta(z)^{-1}
\tau_z\left(\xi(x^{-1},x)^*W(x^{-1})f\right)(e)^*\tau_zg\cr
&=&\int_X\hbox{ d}z\ \Delta(z)^{-1}\left(\sigma_z(\xi(x^{-1},x)^*)
\tau_z(W(x^{-1})f)(e)\right)^*\tau_zg\cr
&=&\int_X\hbox{ d}z\ \Delta(z)^{-1}
\tau_z(W(x^{-1})f)(e)^*\sigma_z(\xi(x^{-1},x))\tau_zg\cr
&=&\int_X\hbox{ d}z\ \Delta(z)^{-1}
\tau_z\tau_{x^{-1}}f(e)^*\xi(z,x^{-1})\xi(zx^{-1},x)\tau_zg\cr
&=&\int_X\hbox{ d}z\ \Delta(z)^{-1}
\tau_{zx^{-1}}f(e)^*\xi(zx^{-1},x)\tau_zg\cr
&=&\int_X\hbox{ d}z\ \Delta(z)^{-1}
\tau_{z}f(e)^*\xi(z,x)\tau_{zx}g\cr
&=&\int_X\hbox{ d}z\ f^*(z)\tau_z\tau_xg\cr
&=&f^*(W(x)g)
\end {eqnarray}
Hence W(x) belongs to the multiplier algebra of ${\cal A}\times_\tau X$
and one has
\begin {equation}
W(x)^*=\xi(x^{-1},x)^*W(x^{-1}),
\label {Wstar}
\qquad x\in X
\end {equation}
\qed
\end {proof}

\begin {proposition}
\begin {equation}
W(x)^*W(x)=W(x)W(x)^*=1,
\qquad x\in X
\end {equation}
\end {proposition}

\begin {proof}
$W(x)^*W(x)=1$ follows from (\ref {isotau}).
In order to prove $W(x)W(x)^*=1$ calculate, using (\ref {Wstar}),
\begin {eqnarray}
& &(W(x)^*f)^*(W(x)^*g)\cr
&=&\int_X\hbox{ d}z\ (\xi(x^{-1},x)^*W(x^{-1})f)^*(z)
\tau_z(\xi(x^{-1},x)^*W(x^{-1})g)\cr
&=&\int_X\hbox{ d}z\ \Delta(z)^{-1}\tau_z(\xi(x^{-1},x)^*\tau_{x^{-1}}f)(e)^*
\sigma_z(\xi(x^{-1},x)^*)\tau_z\tau_{x^{-1}}g\cr
&=&\int_X\hbox{ d}z\ \Delta(z)^{-1}\tau_{zx^{-1}}f(e)^*\xi(z,x^{-1})^*
\sigma_z(\xi(x^{-1},x))\cr
& &\times \sigma_z(\xi(x^{-1},x)^*)\xi(z,x^{-1})\tau_{zx^{-1}}g\cr
&=&\int_X\hbox{ d}z\ \Delta(z)^{-1}\tau_{zx^{-1}}f(e)^*\tau_{zx^{-1}}g\cr
&=&\int_X\hbox{ d}z\ \Delta(z)^{-1}\tau_{z}f(e)^*\tau_{z}g\cr
&=&\int_X\hbox{ d}z\ f^*(z)\tau_{z}g\cr
&=&f^*g
\end {eqnarray}
\qed
\end {proof}

The remainder of the proof of theorem \ref {weylop} is straightforward.
Equation (\ref {sigmaW}) follows immediately from (\ref {zetacon}) and
the unitarity of $W(x)$. Equation (\ref {wlaw}) follows from (\ref {tautau}).

%%%%%%%%%%%%%%%%%%%%%%%%%%%%%%%%%%%%%%%%%%%%%%%%%%%%%%%%%%%%%%%%%%%%%%%%%%%
\section {Application to Quantum Spacetime}

Let ${\bf M}={\bf R}^4,+$ denote Minkowski space and
consider it as a locally compact group.
Let $\Sigma$ be the space of
anti-symmetric 4-by-4 real matrices
of the form
\begin {equation}
\epsilon(e,m)=\left(\matrix{
0    &e_1  &e_2  &e_3 \cr
-e_1 &0    &m_3  &-m_2\cr
-e_2 &-m_3 &0    &m_1 \cr
-e_3 &m_2  &-m_1 &0\cr
}\right)
\end {equation}
satisfying $|e|^2=|m|^2$ and $e.m=\pm 1$.
It is locally compact for the topology induced by the
supremum norm of linear transformations of ${\bf M}$.
A motivation for this particular choice of $\Sigma$
is given in \cite {DFR94}, \cite {DFR95}.
Let ${\cal C}_0(\Sigma,{\bf C})$
denote the $C^*$-algebra of complex continuous functions of
$\Sigma$ vanishing at infinity. A bicharacter
$\xi:\ {\bf M}\times {\bf M}\rightarrow M({\cal C}_0(\Sigma,{\bf C}))$
is defined by
\begin {equation}
\xi(k,k')(\epsilon)=\exp\left({i\over 2}\sum_{\mu,\nu=0}^3 k_\mu
\epsilon_{\mu,\nu}k'_\nu\right),
\qquad k,k'\in {\bf M},\epsilon\in \Sigma
\label {examplexi}
\end {equation}

It is easy to verify that $\xi$ is a $C^*$-multiplier
of the group ${\bf M}$, with
values in ${\cal A}={\cal C}_0(\Sigma,{\bf C})$, and
with associated action $\sigma$ which is trivial,
i.e.~$\sigma_ka=a$ for all $k\in {\bf M}$ and $a \in {\cal A}$.
The crossed product algebra ${\cal A}\times_\tau {\bf M}$,
with $\tau$ the projective action associated with $\xi$,
is an alternative for the algebra constructed in \cite {DFR95}.

Let $\omega$ be a state of ${\cal A}\times_\tau {\bf M}$ for which the maps 
$\lambda\in {\bf R}\rightarrow\omega(g^*\tau_{\lambda k}f)$ are continuous for 
all $f,g\in {\cal A}\times_\tau {\bf M}$ and $k\in {\bf M}$
(this is a so-called {\sl Weyl-state}).
Let $(\pi,{\cal H},\Omega)$ be the GNS-representation induced by 
$\omega$. Then for any $k\in {\bf M}$,
the map $\lambda\in{\bf R}\rightarrow \pi(W(\lambda k))$ is a 
strongly continuous one-parameter group of unitary operators. Hence, by Stone's 
theorem, there exists a self-adjoint operator $Q(k)$ which is the generator of 
this group. Let $e_\mu,\mu=0,1,2,3$ be unit vectors of ${\bf R}^4$. Let 
$Q_\mu\equiv Q(e_\mu)$. On a dense domain one has
$Q(k)=\sum_{\mu=0}^3k_\mu Q_\mu$. Now calculate
\begin {eqnarray}
\exp(ik_\mu Q_\mu)\exp(ik_\nu Q_\nu) 
&=&\pi(W(k_\mu e_\mu)W(k_\nu e_\nu))\cr
&=&\pi(\xi(k_\mu e_\mu,k_\nu e_\nu)W(k_\mu e_\mu+k_\nu e_\nu))\cr
&=&\pi(\xi(k_\mu e_\mu,k_\nu e_\nu))\exp(i(k_\mu Q_\mu+k_\nu Q_\nu)
\label {comlaw}
\end {eqnarray}
Using Stone's theorem one shows that there exist self-adjoint operators 
$R_{\mu,\nu}$ satisfying
\begin {equation}
\pi(\xi(k_\mu e_\mu,k_\nu e_\nu)) 
=e^{ik_\mu k_\nu R_{\mu,\nu}}
\quad\hbox{ and }
R_{\mu,\nu}=-R_{\nu,\mu} 
\end {equation}
Comparing (\ref {comlaw}) with
\begin {equation}
e^{i(A+B)}=e^{iA}e^{iB}\exp({1\over 2}[A,B]), 
\qquad [B,[A,B]]=0
\end {equation}
one concludes that on a dense domain one has 
\begin {equation}
i\big[Q_\mu,Q_\nu\big]=2R_{\mu,\nu}
\quad\hbox{ and }
\big [Q_\mu,R_{\nu,\rho}\big]=0
\label {qcommut}
\end {equation}
The $Q_\mu$ can be 
interpreted as the component operators of the position of a relativistic 
particle. They do not commute, but rather satisfy the commutation relations 
(\ref {qcommut}), which have been proposed in \cite {DFR94,DFR95} as a 
simplified model for quantum spacetime.

The example is rather simple in that both the group ${\bf M}$
and the $C^*$-algebra $\cal A$ are commutative.
The present theory allows for non-commutative groups and algebras.
Hence there are many possibilities for generalization. The 
example shows further that it is necessary to allow for $C^*$-algebras without 
unit. Indeed, the continuity of $k'\rightarrow f\xi(k,k')$ is only true for 
functions $f$ vanishing at infinity. Note that the Weyl-operators $W(k)$ have 
been introduced here taking into account the usual topology of ${\bf M}$,
instead of the discrete topology.
Common complaints against the algebra of the CCR are the necessity
of considering 
the discrete topology on the underlying symplectic space, and the fact that
relevant physical observables do not belong to it -- see e.g.~\cite {LNP98},
IV.3.7 and the notes of IV.3.5.
An alternative for the algebra of the CCR is the crossed product
algebra. It takes the proper topology into account, and contains the
Weyl-operators in its multiplier algebra.

\begin{acknowledgement}
I wish to thank  N.P. Landsman for a helpful discussion
and M. Kuna for critically reading the manuscript.
\end{acknowledgement}

%%%%%%%%%%%%%%%%%%%%%%%%%%%%%%%%%%%%%%%%%%%%%%%%%%%%%%%%%%%%%%%%%%%%%%%%%%%

\end {document}